\documentclass[a4paper,11pt]{article}
\usepackage{jcappub} 

\usepackage[parfill]{parskip}
\usepackage{graphicx}
\usepackage{amssymb}
\usepackage{amsmath}
\usepackage{mathtools}
\usepackage{tikz}
\usepackage{color}
\usepackage{caption}
\usepackage{ulem}
\usepackage{enumitem}

\usetikzlibrary{shapes.geometric, arrows}

\tikzstyle{io} = [trapezium, 
trapezium stretches=true, 
trapezium left angle=70, 
trapezium right angle=110, 
minimum width=3cm, 
minimum height=1cm, text centered, 
text width=3.7cm,
draw=black, fill=orange!30]

\tikzstyle{startstop} = [rectangle, rounded corners, 
minimum width=3cm, 
minimum height=1cm,
text centered, 
text width=3.7cm,
draw=black, 
fill=orange!30]

\tikzstyle{decision} = [diamond, 
minimum width=1cm, 
minimum height=0.5cm, 
text centered, 
text width=1.6cm,
draw=black, 
fill=orange!30]

\tikzstyle{arrow} = [thick,->,>=stealth]

\newcommand{\be}{\begin{equation}}
\newcommand{\ee}{\end{equation}}
\newcommand{\ba}{\begin{eqnarray}}
\newcommand{\ea}{\end{eqnarray}}

\begin{document}

\title{Testing gravity with the latent heat\\ of neutron star matter}

\author{Pablo Navarro Moreno$^1$, Aneta Wojnar$^{1,2}$ \\ and Felipe J. Llanes-Estrada$^{1,3}$}
\affiliation{$^1$ Departamento de F\'isica Te\'orica \& IPARCOS, Universidad Complutense de Madrid,  \\ E-28040, 
Madrid, Spain\\
$^2$ Institute of Theoretical physics, University of Wroc\l aw, pl. Maxa Borna 9, 50-206 Wroc\l aw, Poland\\
$^{3}$ On leave at the Theory Department of CERN, 1211 Geneva, Switzerland.}

\emailAdd{panava03@ucm.es}

\abstract{
The Seidov limit is a bound on the maximum latent heat that a presumed first-order phase transition of neutron-star matter can have before its excess energy density, not compensated by additional pressure, results in gravitational collapse. Because latent heat forces an apparent nonanalytic behaviour
in plots correlating physical quantities (kinks in two-dimensional, ridges in three-dimensional ones), it can be constrained by data.
As the onset of collapse depends on the intensity of gravity, testing for sudden derivative changes and, if they are large, breaching the Seidov limit would reward with two successive discoveries: such a phase transition (which could stem from hadron matter but also from a gravitational phase transition), and a
modification of General Relativity (thus breaking the matter/gravity degeneracy). We illustrate the point with 
$f(R)=R+\alpha R^2$ metric gravity. 
}

\maketitle

\section{Introduction}
\label{sec:Intro}

General Relativity (GR), proposed by A. Einstein in 1915, remains the widely accepted theory of gravity and has undergone extensive testing in the weak field regime. Numerous astrophysical observations, including solar system tests, binary pulsars, and gravitational-wave phenomena \cite{LIGOScientific:2020kqk, LIGOScientific:2018hze}, have consistently supported the predictions of GR. Nevertheless, in the strong field regime, GR may exhibit limitations or fail to accurately describe certain phenomena. Therefore, it is wise to continue testing GR against relatively straightforward alternative theories in these new regimes.

Neutron stars (NS) are among the most compact, non-collapsed objects observable, making them excellent laboratories for testing various theories of gravity~\cite{Hendi:2015pua,Hendi:2017ibm,AparicioResco:2016xcm}. In particular, their high density allows us to test the large stress-energy tensor regime. Studying NS in the context of modified gravity theories can help constrain the parameters of these theories. Observations from X-ray emissions, binary radio systems, and gravitational waves provide constraints on the mass and radius of these astrophysical objects. Table \ref{table1} presents some observational {masses and angular frequencies from \cite{Ozel:2016oaf,Antoniadis:2016hxz,Salmi:2024aum}.
Radii are not yet so directly extracted, but NICER has determined, for example, that the radius of J0740+6620 is about $12.5^{+1.3}_{-0.9}$ km.
}

\begin{table}[h!]
\centering
\begin{tabular}{|c|c|c|}
\hline
Name &  $\Omega({\rm rad}\cdot {\rm ms}^{-1})$ & $M(M_{\odot})$ \\ \hline
J0337+1715 & 2.299 & 1.4401(15) \\ \hline
J0348+0432 & 0.161 & 2.01(4) \\ \hline
J0509+380 & 0.082 & 1.34(8) \\ \hline
J0453+1559 & 0.137 & 1.559(5) \\ \hline 
J0740+6620 & 2.177 & 2.07(7)\\ 
\hline
J1012+5307 & 1.195 & 1.72(16) \\ \hline
\end{tabular}
\caption{Angular velocity and mass of a few well-measured pulsars \cite{Ozel:2016oaf,Antoniadis:2016hxz,Salmi:2024aum}.}
\label{table1}
\end{table}

NS masses typically range from 1 to 2 $M_{\odot}$, with radii spanning 10 to 13 km. The primary challenge in studying NS lies in the uncertainty of the Equation of State (EoS) that accurately describes the matter within them. This uncertainty complicates efforts to constrain the parameters of gravitational theories. To describe NS within the framework of modified gravity, it is essential to employ EoS that are independent of astrophysical observations and are instead constrained solely by microscopic physics \cite{Lope-Oter:2023urz,Lope-Oter:2021vxl}.

The most popular modified gravity theories nowadays are $f(R)$ and scalar-tensor theories. $f(R)$ gravity is a natural generalization of GR in which the Ricci scalar $R$ in the action is replaced by a more general function of it. By construction, these theories introduce dimensionful parameters which must be constrained by observations. $f(R)$ theories are a particular case of scalar-tensor theories, which include both a tensor field and a scalar field to mediate the gravitational interaction \cite{Yazadjiev:2014cza,Staykov:2016mbt}. Scalar-tensor theories become important in inflationary cosmology \cite{Starobinsky:1980te}.

The aim of this project is to establish the Seidov limit for the latent heat in a phase transition as an additional diagnostic which may establish the need for corrections to General Relativity.
For this we study static neutron stars within alternative theories of gravity. As an illustrative example  we adopt $R^2$-gravity, with an additional parameter $\alpha$ in which 
\begin{equation}\label{deffofR}
  f(R)=R+\alpha R^2\ ,
\end{equation} 
but we point out that the static Tolman-Oppenheimer-Volkoff (TOV) equations which we will present are more general and valid in a broader family of theories which can be characterized by presenting a change of the intensity of gravity and a geometric shift of the Einstein tensor as can be seen in Eq.~(\ref{eq.campo}).
We study both the static (computing the typical mass-radius diagrams for different families of stars and analyzing the maximum latent heat that a star can support before collapsing) and the rotating star \cite{Staykov:2014mwa,Doneva:2016xmf} (computing the moment of inertia and other observables) for different EoS and values of the parameter $\alpha$ of the theory  in Eq.~(\ref{deffofR}). As a check, we compare the results in the $\alpha\to 0$ limit of General Relativity with our previous work~\cite{Moreno:2023xez}, finding excellent agreement. In what follows, we will consider only positive values of the parameter $\alpha$. In the case of its negative branch, the gravitational mass receives large contributions from supposedly empty space outside the star.
This makes the negative branch {difficult to interpret} \cite{AparicioResco:2016xcm}. No stable configurations are found for certain specific EoS \cite{Astashenok:2018iav}.
Conversely, for $\alpha >0$, a match with the asymptotically Schwarzschild solution is feasible. For more discussion of $f(R)$ models in NS physics, see \cite{Olmo:2019flu}.

The article is concerned with static, equilibrated neutron stars, in which the phase transition has already taken place. The latent heat is then manifest as some nonderivability in the thermodynamic equilibrium quantities which may leave an effect. Cataclysmic effects should happen upon the core-collapse supernova process giving birth to the neutron star, and have long dissipated by the time our treatment to a cold star applies.

We believe that our work establishes a) the measurability of latent heat through nonanalyticities (sudden derivative changes) in physical plots and b) the possibility to employ the Seidov limit to push beyond GR, just as breaking the largest-mass ceiling would~\cite{Astashenok:2021peo}.
This can happen because gravity's attraction (which ultimately causes collapse) can be weakened in modified theories~\cite{Albareti:2012va}.

The article is organized as follows. In Section~\ref{sec:modgrav} we recall a family of modified gravity theories characterized by a scalar weakening/strengthening factor and a geometric shift. 
A subclass of that family (subsection~\ref{sec:f(R)}) is that of $f(R)$ theories, in particular $R-$squared gravity, and their equivalence (subsection~\ref{subsec:equivalencetoscalartensor}) to particular scalar-tensor theories in the \textit{Einstein frame}. 

In section~\ref{sec:GeneralFormalism} we then discuss the TOV equations of hydrostatic equilibrium {for $f(R)$ modified gravity, including boundary conditions and initial conditions for the radial integration, which require a bit more care than in GR. } In Section~\ref{sec:matter} we then turn to the matter content of the neutron star. The EoS uncertainty band which we employ is discussed in subsection~\ref{subsec:EoS}. 
Subsection~\ref{sec:Latent-Heat} is then dedicated to defining the latent heat for a phase transition in a given EoS to an exotic phase of hadron matter.

Section~\ref{sec:Seidov-limit} is then dedicated to repeating Seidov's reasoning for the maximum latent heat that a star can support before collapse (which is a bound to the maximum latent heat which could be measured in a static neutron star, even if hadron physics would allow for larger ones), and importantly, we extend the calculation to $R+\alpha R^2$ gravity. 
{Section~\ref{sec:Bouchdahl-limit} is dedicated to the Buchdahl-Bondi limit in the mass-radius diagram for $f(R)$ theory, another known result which might be of some use in the attempt to distinguish matter from gravity effects in neutron stars. 
}
Then, in Section \ref{sec:Rotating-star} we discuss the slowly rotating star approximation, {just to show that ridges/kinks
in physical properties due to phase transitions
are by no means reduced to the mass-radius diagram}.

{Although the original Seidov limit (which we reobtained also for $f(R)$ gravity) was formulated in the small-core approximation, we turn to numerical computations to make it more generally applicable}. 
The field equations obtained earlier in subsection~\ref{sec:field-eqs} 
are then numerically solved, in Section~\ref{sec:numerics}, to illustrate the points made, explaining how the numerical algorithm is designed.

{Finally, Section~\ref{sec:Conclusions}, recapitulates the discussion and concluding remarks as well as ongoing investigations are mentioned. 
}

\section{A simple class of modified gravity theories}\label{sec:modgrav}

There is a quite generic family of modified gravity theories whose field equations can be written as \cite{Wojnar:2016bzk}
\begin{equation}\label{eq.campo}
    \sigma(\chi)(G_{\mu\nu}-W_{\mu\nu})=\kappa T_{\mu\nu},
\end{equation}
where $G_{\mu\nu}=R_{\mu\nu}-\frac{1}{2}Rg_{\mu\nu}$ is the Einstein tensor, $T_{\mu\nu}$ the stress-energy one ($\kappa=8\pi$ as we work in geometrized units $G=c=1$) and $\sigma(\chi)$ is the coupling to the gravitational field (due to other fields or gravitational curvature invariants, generically denoted by $\chi$), which acts as a gravitational weakening/strengthening parameter. $W_{\mu\nu}$ is a symmetric tensor that may include additional terms depending on the theory considered and it shifts the geometrical contribution of the theory away from GR.  Notice that we recover the general relativistic field equations taking $\sigma(\chi)=1$ and $W_{\mu\nu}=0$.

We consider, as the {$0^{\rm th}$ order approximation to matter}, the energy-momentum tensor of a perfect fluid,  $T_{\mu\nu}=(\rho+p)u_{\mu}u_{\nu}+pg_{\mu\nu}$, where $u_{\mu}$ is the 4-velocity (which satisfies $u^2=-1$) of an observer moving with the fluid. 

In these theories of modified gravity it is generally not the canonical energy-momentum tensor $T_{\mu\nu}$ which is conserved. Instead, consistency with the Bianchi identity requires the conservation of a rescaled, shifted tensor $T_{\mu\nu}^{\rm{eff}}=\frac{1}{\sigma(\chi)}T_{\mu\nu}+\frac{1}{\kappa}W_{\mu\nu}$.

One salient and well-known class of theories which can be framed as in Eq.~(\ref{eq.campo}) is that of $f(R)$ metric gravity, and we now turn to it.

\subsection{$f(R)$ theories }
\label{sec:f(R)}
These theories can be cast in the form of Eq.~\eqref{eq.campo} with $\sigma =f'(R)$ and an adequate $W_{\mu\nu}$, {as apparent in Eq.~(\ref{field.eqs.fR}) below.} 
They are built modifying the Einstein-Hilbert action of general relativity ~\cite{Yazadjiev:2014cza,Bhattacharyya:2022dzh}, replacing the scalar curvature $R$ with a function of the same that introduces additional parameters in the theory. There exist different formalisms for $f(R)$ theories, such as the metric formalism or the Palatini one, in which the metric tensor $g$ and the connection $\Gamma$ are independent variables \cite{Lope-Oter:2023urz} (to see a general review of metric and Palatini theories in application to stellar objects, see \cite{Olmo:2019flu}). Here, we focus on the metric one. The action is given by
\begin{equation}\label{fR.action}
    S=\frac{1}{2\kappa}\int d^4x\sqrt{-g}f(R)+S_M(g_{\mu\nu},\chi),
\end{equation}

where $S_M$ is the action of the matter fields $\chi$. A theory formulated over a stable vacuum must obey $\frac{d^2 f}{dR^2}\geq 0$ and $\frac{df}{dR}>0$ \cite{Yazadjiev:2014cza}. Varying the action with respect to the metric tensor we obtain the known field equations

\begin{equation}\label{field.eqs.fR}
    f'(R)R_{\mu\nu}-\frac{1}{2}f(R)g_{\mu\nu}-\left[\nabla_{\mu}\nabla_{\nu}-g_{\mu\nu}\Box\right]f'(R)=\kappa T_{\mu\nu}.
\end{equation}

For the linear  $f(R)=R$ we recover the Einstein field equations of general relativity.

\subsection{Equivalence to scalar-tensor theories}\label{subsec:equivalencetoscalartensor}
$f(R)$ theories are equivalent to Brans-Dicke scalar-tensor theories 
{(with vanishing Brans-Dicke coupling constant $\omega_{BD}=0$) and a non-zero potential of the scalar field:}
\begin{equation}\label{S_st}
    S=\frac{1}{2\kappa}\int d^4x\sqrt{-g}\left(\varphi R-U(\varphi)\right)+S_M(g_{\mu\nu},\chi),
\end{equation}
with the scalar field defined as $\varphi\equiv f'(\psi)$, its potential being
\begin{equation}
U(\varphi)\equiv\psi(\varphi)f'(\psi(\varphi))-f(\psi(\varphi)) ,
\end{equation}
and where the argument $\chi$ of $S_M$ includes any other matter fields
 Note that $'$ denotes here derivative with respect to a new field $\psi$ and one imposes $f''(\psi)\neq 0$ to get $R=\psi$.

In particular, we can work with a quadratic $f(R)$ theory with the functional form $f(R)=R+\alpha R^2$ \cite{Starobinsky:1980te}, where $\alpha > 0$, to obtain nominally stable solutions \cite{Yazadjiev:2014cza}. From the definitions of $\varphi$ and $U(\varphi)$ we then get
\begin{equation}\label{Upotential}
    U(\varphi)=\frac{1}{4\alpha}(\varphi-1)^2.
\end{equation}
The field equations and the equation of the motion of $\varphi$ can be obtained by varying the action (\ref{S_st}) with respect to $g_{\mu\nu}$ and $\varphi$ respectively \cite{Sotiriou:2008rp}, leading to 
\begin{equation}\label{eq.varphi}
    G_{\mu\nu}=\frac{\kappa}{\varphi}T_{\mu\nu}-\frac{1}{2\varphi}g_{\mu\nu}U(\varphi)+\frac{1}{\varphi}(\nabla_{\mu}\nabla_{\nu}\varphi-g_{\mu\nu}\Box\varphi),
\end{equation}
\begin{equation}\label{eq.varphi2}
    3\Box\varphi+2U(\varphi)-\varphi\frac{dU}{d\varphi}=\kappa T.
\end{equation}

In order to simplify the further studies, one can rewrite the above equations in the so-called \textit{Einstein frame} (EF). They are related to the \textit{Jordan frame} ones through a conformal transformation\footnote{Notice that the conformal transformation might be singular in some particular cases. We discuss this problem in the further part of the paper.}, with the conformal factor defined as ${\mathcal{D}}^2(\phi)\! =\! \varphi\!=\!e^{\frac{2\phi}{\sqrt{3}}}$, that rescales the EF metric 
\begin{equation}\label{conformal_transf}
    g^*_{\mu\nu}={\mathcal{D}}^{2}(\phi)g_{\mu\nu},
\end{equation}
where we have redefined the scalar field as
\begin{equation}\label{eq.phivarphi}
    \phi\equiv \frac{\sqrt{3}}{2}\log\varphi.
\end{equation}
We will denote geometric quantities in the \textit{Einstein frame} {by a raised asterisk} *. 
From Eq.~(\ref{S_st}) the action is then
\begin{equation}\label{S_st2}
    S=\frac{1}{2\kappa}\int d^4x\sqrt{-g^*}\left(R^*-2g^{*\mu\nu}\nabla^*_{\mu}\phi\nabla^*_{\nu}\phi-\right.
    \left.-V(\phi)\right)+S_M({\mathcal{D}}^{-2}g^*_{\mu\nu},\chi),
\end{equation}
where the potential takes the form
\begin{equation}
    V(\phi)=\frac{1}{4\alpha}\left(1-e^{-\frac{2}{\sqrt{3}}\phi}\right)^2.
\end{equation}
Before computing the field equations it is convenient to check what happens to the physical magnitudes, such as the pressure and energy density, whenever a conformal transformation is performed \cite{dabrowski2009conformal}. Notice that $\rho$ and $p$ are the physical energy density and pressure, the ones which enter the equation of state due to microscopic physics. 
To the energy momentum tensor which they produce we apply the conformal transformation (\ref{conformal_transf}), $T^*_{\mu\nu}={\mathcal{D}}^{-2}T_{\mu\nu}$. Raising both indices we get $T^{*\mu\nu}={\mathcal{D}}^{-6}T^{\mu\nu}$, and its trace is conformally transformed to $T^*={\mathcal{D}}^{-4}T$. From the definition of the energy-momentum tensor for a perfect fluid immediately follows that $\rho^*={\mathcal{D}}^{-4}\rho$ and $p^*={\mathcal{D}}^{-4}p$. Lastly, the 4-velocity of the comoving observer with the fluid is transformed as $u^*_{\mu}={\mathcal{D}} u_{\mu}$ and $u^{*\mu}={\mathcal{D}}^{-1}u^{\mu}$.

Keeping this in mind, the field equations are now (see e.g. \cite{Yazadjiev:2014cza}):
\begin{equation}\label{eq.EFG}
    G^*_{\mu\nu}=\kappa T^{*(M)}_{\mu\nu}+2\partial_{\mu}\phi\partial_{\nu}\phi-g^*_{\mu\nu}\partial^{\rho}\phi\partial_{\rho}\phi 
    -\frac{1}{2}g^{*}_{\mu\nu}V(\phi)=\kappa T^{*(M)}_{\mu\nu}+2T^{*(\phi)}_{\mu\nu},
\end{equation}
\begin{equation}\label{eq.phi2}
    \Box^*\phi-\frac{1}{4}\frac{dV}{d\phi}=\frac{\kappa}{2\sqrt{3}} T^*,
\end{equation}
where $\Box^*=\nabla_{\mu}^*\nabla^{*\mu}$. 

In Sec.~\ref{sec:field-eqs}, we are going to solve the static star computing the field equations in the \textit{Einstein frame}. Thus, it is convenient to compute the divergence of the energy-momentum tensor therein. From the Bianchi identities, using (\ref{eq.phi2}) and taking into account that the connection is symmetric, we arrive at
\begin{equation}
    \nabla_{\mu}^*T^{*\mu}_{\nu}=-\frac{1}{\sqrt{3}}T^*\nabla^*_{\nu}\phi.
\end{equation}

\section{Static stars in modified gravity}
\label{sec:GeneralFormalism}
\subsection{Generalized TOV equations for  $f(R)$}\label{sec:field-eqs}

We may now address neutron stars in this theory following many earlier studies, {\it e.g.}~\cite{Olmo:2019flu} for a review. 
We initially formulate the problem in the \textit{Einstein frame} in which the equations take a form closest to that of the general relativistic TOV system, starting again from the static and spherically symmetric metric
\begin{equation}\label{metrica}
    ds^2=-e^{2\nu(r)}dt^2+e^{2\lambda(r)}dr^2+r^2(d\theta^2+\sin^2{\theta}d\phi^2),
\end{equation}
but now for $ds_*^2$. (Once the field equations (\ref{eq.EFG}) and (\ref{eq.phi2}) are obtained, the conformal transformation is inverted, returning the components of the metric to the \textit{Jordan frame}.) The components $G^*_{tt}$ and $G^*_{rr}$ together with the divergence of $T^*_{\mu\nu}$ immediately
yield the modified-gravity TOV equations for $f(R)$, formulated as a scalar-tensor theory, 
\begin{equation}\label{EF_Gtt}
    \frac{1}{r^2}\frac{d}{dr}\left[r\left(1-e^{-2\lambda}\right)\right]=\frac{8\pi}{{\mathcal{D}}^{4}}\rho+e^{-2\lambda}\phi'^2+\frac{1}{2}V(\phi),
\end{equation}  
\begin{equation}\label{EF_Grr}
\frac{2}{r}e^{-2\lambda}\frac{d\nu}{dr}+\frac{\left(e^{-2\lambda}-1\right)}{r^2}=\frac{8\pi}{{\mathcal{D}}^{4}}p+e^{-2\lambda}\phi'^2-\frac{1}{2}V(\phi),
\end{equation}
\begin{equation}\label{EF_Tmur}
    \frac{dp}{dr}=-(\rho+p)\left[\frac{d\nu}{dr}-\frac{1}{\sqrt{3}}\frac{d\phi}{dr}\right].
\end{equation}

Equation (\ref{eq.phi2}) is explicitly written as 
\begin{equation}\label{eq.phi22}
    \phi''+\left[\frac{d\nu}{dr}-\frac{d\lambda}{dr}+\frac{2}{r}\right]\phi'=\frac{\kappa}{2\sqrt{3}}e^{2\lambda}{\mathcal{D}}^{-4}(3p-\rho)+\frac{1}{4}e^{2\lambda}\frac{dV}{d\phi}.
\end{equation}
Notice that $\rho$ and $p$ are the physical energy density and pressure in the \textit{Jordan frame} respectively, related by the equation of state (EoS). Thus, the system of differential equations (\ref{EF_Gtt})-(\ref{eq.phi22}) together with a given EoS, completely determine the problem of the static star in this modified theory of gravity. However, before we move to the numerical analysis, let us discuss a possible singular behaviour of the above equations, caused by the conformal factor $\mathcal D$ when its values vanishes. This characteristic arises from the fact that {gravity} modifications often introduce new matter-dependent contributions to the hydrostatic equilibrium equation through the modified Klein-Gordon equation \eqref{eq.phi22}, which links the dynamics of the scalar field with ordinary matter sources - here, by the value of the trace of the energy-momentum tensor. Additionally, it is important to note that the critical value of the parameter {$\alpha$ at which this happens} varies with the energy density and pressure, depending on their profiles. Therefore, careful consideration must be given when selecting a specific value for $\alpha$, taking into account the equation of state and the scalar-tensor model in question. In the following analysis, we {continue focusing} on a quadratic model, specifically $f(R) = R + \alpha R^2$, {so that $\mathcal{D}^2=f'=1+2\alpha R$}.

\begin{figure}[h!]
\centering
\includegraphics[width=\textwidth]{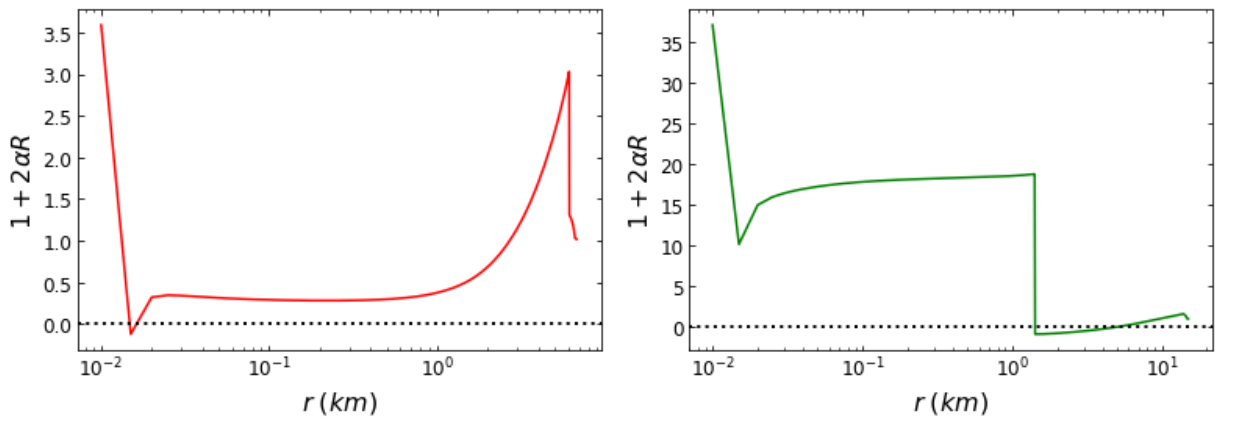}
\caption{
Dilatation factor $(1+2\alpha R)$ inside a neutron star. 
Left: for an extremely soft EoS (ExS given below in Fig.~\ref{fig:EoS}) with central pressure $p_c=1000\;\rm{MeV/fm^3}$ and $\alpha=25$km$^2$. The transformation between the Einstein and the Jordan frame is singular when the solid line touches the dotted line at zero. Right: same but for the stiffest EoS ExR, in this case with modified-gravity parameter $\alpha=50$km$^2$.}
\label{fig:R}
\end{figure}

Fig.~\ref{fig:R} presents our numerical computation of that dilatation factor  (basically, the Ricci scalar) in the interior of two stars characterized by extremely soft (smallest pressure at given density) and extremely hard (largest pressure at each density) equations of state (see subsection~\ref{subsec:EoS}), respectively. As seen in the figure, the factor $\mathcal{D}^2$ vanishes at some point inside each of the stars, meaning that the joint theory defined by $G=1$, the given $\alpha$ and the respective EoS in that graph is inconsistent with that central pressure.

{For example, adopting the stiffest EoS ExR, and a central  $p_c=4000$MeV/fm$^3$, the ``critical'' value (for which $\mathcal{D}^2$ becomes negative at some point inside the star) is found to be around $\alpha=20$km$^2$.
For lower central pressure, the minimum value of $\alpha$ at which $\mathcal{D}^2$ crosses zero rises. Conversely, for softer EoS, it diminishes.}

{However, this does not necessarily mean that every value of $\alpha$ is excluded, but that the stars which can be supported depend on the $\alpha$.
For every EoS and every $\alpha$, there always are stars in which $\mathcal{D}^2\neq 0$ everywhere. 
If $\alpha$ is increased, the size  of the allowed $p_c$ decreases, but that is all.}

{
It then becomes an observational question: is the slate of physically allowed configurations for each $(\alpha,{\rm EoS})$ combination  sufficient to explain the statistics of observed neutron stars? 
}
\subsection{Initial and boundary conditions for the radial integration}

The initial conditions and boundary conditions that we will impose are the following, naturally arising in the \textit{Einstein frame}. Pressure in the center of the star is a given value $p(0)=p_c$. Integrating the system for different central pressures will produce a family of solutions. The radius $R$ of the star will be determined by the condition $p(R)=0$. The star's physical radius  will then be 
\begin{equation}\label{R_star}
    R_s=R{\mathcal{D}}^{-1}(\phi(R)).
\end{equation}
The condition $\lambda(0)=0$ guarantees regularity of the metric and on the other hand, $\displaystyle \frac{d\phi}{dr}(0)=0$ that of the scalar field. We will also impose our spacetime to be asymptotically flat so that $\displaystyle \lim_{r\to\infty}V(\phi)=0$ and then $\displaystyle \lim_{r\to\infty}\phi=0$. Moreover $\displaystyle \lim_{r\to\infty}\nu(r)=0$. These guarantee asymptotic flatness in both \textit{Einstein} and \textit{Jordan frames}.

\textbf{Note that in the presence of matter, the frames are not equivalent due to the coupling term with matter that appears in the \textit{Einstein frame}. Similar to the radius \eqref{R_star}, the values of other physical quantities, such as pressure $p$ and density $\rho$, which are related to each other by microphysical descriptions, are taken in the \textit{Jordan frame}. It is common practice to use equations in the Einstein frame because they are simpler compared to those in the Jordan frame (and also because most stellar codes are developed for GR). However, the physical quantities are considered in terms of the Jordan frame values. Note that in equation \eqref{EF_Gtt} and subsequent ones, there is a function $\mathcal{D}$ associated with the physical fields. For example, $\rho_E=\rho/\mathcal{D}^4$ represents the "\textit{Einstein frame} density" (see the relations between frames in \cite{dabrowski2009conformal}).
}

Due to the form of the conformal factor and to the fact that the scalar field exponentially decreases at infinity, we find that the masses of the star in both \textit{Einstein} and \textit{Jordan frames} coincide. In order to obtain their common value we only need to compare the exterior metric towards infinity with that given by the Schwarzschild metric, so that
\begin{equation}\label{lim_asint}
    e^{2\lambda(r)}\longrightarrow \left(1-2\frac{M}{r}\right)^{-1}.
\end{equation}
\textbf{This is possible and simple for $\alpha>0$ as the scalar field dies away with $r$ without oscillation. Note that in metric $f(R)$ gravity, multiple exterior solutions exist, one of which is the Schwarzschild-de Sitter solution, as briefly discussed in \cite{Olmo:2019flu}. This indicates that the usual Birkhoff theorem does not apply; for its generalized version, see \cite{Nzioki:2009av}. This can, in principle, create challenges when matching interior and exterior solutions (see \cite{Multamaki:2006ym,Henttunen:2007bz}), although a specific procedure for this model of gravity has been adapted to address these issues in \cite{Senovilla:2013vra,Casado-Turrion:2023rni}. On the other hand, in the case of not decaying scalar field outside the star (the so-called “gravisphere” \cite{Astashenok:2017dpo}) which can arise to a problem of the well defined mass of the star but at the same time also providing testing tools. Moreover, for the general class of the scalar-tensor theories, one introduces the screening mechanism (see e.g. \cite{Jain:2010ka}}.

This results in the total mass of the star, numerically extracted from the solution to Eq.~(\ref{EF_Tmur}),
\begin{equation}\label{masa}
    M=\lim_{r\to\infty}\frac{r}{2}\left(1-e^{-2\lambda(r)}\right)\ .
\end{equation}

The directly observable physical magnitudes of the static star are its mass and radius given by equations \eqref{R_star} and \eqref{masa}. The system of differential equations \eqref{EF_Gtt}-\eqref{eq.phi22}, together with EoS, completely determines the problem.

We now turn to the actual energy-matter content of the star and a brief discussion about the latent heat which characterises first-order phase transitions thereof.

\section{Matter-energy and phase transitions}\label{sec:matter}
\subsection{Equations of State}
 \label{subsec:EoS}

In this section we present the equations of state employed to solve the field equations. The EoS relate the thermodynamic variables that describe the state of matter under certain physical conditions. Added to the set of equations \eqref{EF_Gtt}-\eqref{eq.phi22}; this completes the modified TOV system. Assuming that the fluid that constitutes the neutron star is a barotropic one, the equation of state takes the form $p=p(\rho)$. 
{Since the thermodynamically exact EoS is never exactly known, uncertainty bands based on hadron input are standard.
}

\begin{figure}[h!]
\centering
\includegraphics[width=8cm]{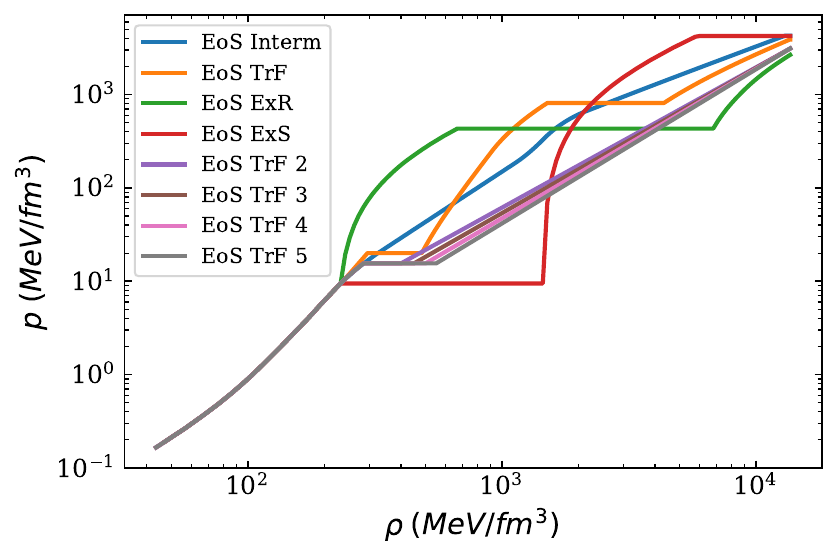}
\caption{Equations of state in logarithmic scale. Pressure ($p$) as a function of the energy density ($\rho$) in $\rm{MeV/fm^3}$. The green line is the stiffest EoS in the low-density neutron star region, while the red line represents the softest allowed EoS there.}
\label{fig:EoS}
\end{figure}

The EoS used here are shown in Fig.~\ref{fig:EoS}: the most rigid (red) and the softest (green) and some typical intermediate case. 

This EoS band, provided by our research group \cite{Lope-Oter:2021vxl}, is valid for cold hadronic matter. It is obtained from Chiral Perturbation Theory for low densities and perturbative Quantum Chromodynamics for high densities. Intermediate densities are obtained interpolating between both branches. Furthermore, this family of EoS is not constrained by any astrophysical observable so it can be used in extensions of general relativity, such as in scalar-tensor or $f(R)$ theories here.

All the EoS of the family satisfy the stability and causality conditions given by $\displaystyle\frac{dp}{d\rho}\geq 0$ and $\displaystyle\frac{dp}{d\rho}\leq 1$, respectively. Additionally to these conditions, the family of EoS also satisfies thermodynamic consistency $\displaystyle p=\int n(\mu)d\mu$ for a causal $n(\mu)$~\cite{Komoltsev:2021jzg}.

In general, a stiff EoS has a large slope in the pressure-energy density diagram that can even saturate causality ($c_s^2\lesssim 1$). The most rigid EoS employed in this work is dubbed EoS ExR (an acronym for Extremely Rigid) and the softest is EoS ExS (likewise abbreviating Extremely Soft). Upon increasing the stellar mass, a larger slope in the EoS will lead to an increase in the radius, a smaller slope to a decreased one. In this work we analyze these extreme cases and some intermediate ones in section \ref{sec:field-eqs}.

Of interest for our main thrust is that some of the EoS in Fig.~\ref{fig:EoS} present first order phase transitions (horizontal straight lines) given by

\begin{equation}\label{phase_trans}
    dp/d\rho=0.
\end{equation}

In the following results we  identify some of the nonanalyticities in NS observables that these phase transitions induce\footnote{Remember that nonanalyticity is necessary for any nonzero function with zero derivative on a finite segment.}.

\subsection{Latent heat}
\label{sec:Latent-Heat}

We now turn to an equation of state with a first order phase transition at $p_1$ from $\rho_1$ to $\rho_2>\rho_1$. First, we consider a relativistic star described by the TOV equations (\ref{EF_Gtt})-  (\ref{EF_Tmur}). If the central pressure exceeds $p_1$, then we obtain a star with a core in a new phase. If it is lower than $p_1$ then we have a phase-homogeneous star.

The intensity of a first order phase transition is quantified by the latent heat, which can be defined by \cite{Lope-Oter:2021mjp}
\begin{equation}\label{Latent_heat}
    L=p_1\frac{\rho_2-\rho_1}{\rho_1\rho_2}.
\end{equation}
This is a natural and practical definition of latent heat in the context of a neutron star. It is close to the naive $dL = dE/(nM)$ from which it differs by the binding (or antibinding) energy per nucleon, B/A, with the difference quantified in our earlier work \cite{Lope-Oter:2021vxl}, whose first section details the derivation. But it is more practical because it is entirely written in terms of energy densities and pressures which directly appear in the stress-energy-momentum tensor, without requiring further theory.

Seidov's study within GR, employing the small-core limit \cite{Seidov} predicts a critical value such that for larger $L$ the star is unstable: if the phase transition is long enough, gravitational collapse occurs. He found the strongest phase transition allowed in GR (for a small core) to have an energy-density jump 
\begin{equation}\label{Latent_heat2}
    \rho_2-\rho_1=\rho_1\left(\frac{1}{2}+\frac{3}{2}\frac{p_1}{\rho_1}\right)
\end{equation}
which our group has formulated as a latent heat in earlier work~\cite{Lope-Oter:2021mjp}.

In the context of modified gravity we can integrate the TOV-like system  for all the equations of state and obtain an approximate limit for the maximum latent heat allowed. 

\section{Seidov limit in $R^2$-gravity}
\label{sec:Seidov-limit}

We now {study the generalization of} Seidov's limit for the particular $R+\alpha R^2$ theory in the Einstein frame. To do so, we first rewrite the TOV system of differential equations. It is convenient to adopt the Schwarzschild notation
\be\label{1}
    e^{2\lambda}=\left(1-\frac{2m}{r}\right)^{-1},
\ee
and so, the TOV system is written as
\be\label{2}
    \frac{dm}{dr}=4\pi r^2\rho{\mathcal{D}}^{-4}+\frac{r}{2}(r-2m)\phi'^2+\frac{r^2}{4}V(\phi),
\ee

\be\label{3}
    \frac{dp}{dr}=-\frac{\rho+p}{r-2m}\left[\frac{m}{r}+\frac{4\pi}{{\mathcal{D}}^{4}}r^2p+(r-2m)\left(\frac{r}{2}-\frac{1}{\sqrt{3}}\right)\phi'^2-\frac{r^2}{4}V(\phi)\right].
\ee

The equation of motion of the scalar field can be rewritten in the following way:
\be\label{scalar_field_EoM}
    \phi''+\frac{r^2}{2(r-2m)}\left[-\frac{4m}{r^3}+\frac{8\pi}{{\mathcal{D}}^4}(p-\rho)-V(\phi)+\frac{4}{r^2}\right]\phi'=\frac{r}{r-2m}\left[\frac{4\pi}{\sqrt{3}{\mathcal{D}}^4}(3p-\rho)+\frac{1}{4}\frac{dV}{d\phi}\right].
\ee

In what follows, we suppose that the central pressure of the star $p_c$ and the pressure $p_1$, at which the phase transition takes place, satisfy $p_c-p_1=\delta$, where $\delta\ll p_1$. At $p=p_1$ there is a discrete change in the energy density {between phase 1 and phase 2} from $\rho_1$ to $\rho_2=q\rho_1 >\rho_1$. If a star's central pressure $p_c$ is greater than $p_1$, then the star presents a nucleus in the new phase 2, whilst if $p_c$ is lower than $p_1$ the star is homogeneous in phase 1. Following Seidov, we call {$p_+(r)$ the homogeneous-star solution with $p_c$ slightly greater than $p_1$ and $p_-(r)$  that with $p_c=p_1$.}
We can then relate both solutions by introducing a perturbation function $\Pi(r)$ as follows,

\be\label{4}
    p_+(r)=p_-(r)+\Pi(r).
\ee

In a small neighbourhood near the center of the star, $p_-(r)$ takes the form

\be\label{5}
    p_-(r)=p_1-\delta_1(r).
\ee

We can expand $\phi(r)$ in power series as $r\longrightarrow 0$, $\displaystyle\phi(r)=\phi_c+\phi_c'r+\frac{\phi_c''}{2!}r^2+\dots$ The initial conditions are satisfied iff $\phi_c'=0$. Thus, we can approximate the scalar field near the origin by its central value $\phi\approx\phi_c$. The next order in the power series can be obtained from \eqref{scalar_field_EoM}, yielding $\displaystyle \frac{\phi_c''}{2!}=\frac{2\pi}{3\sqrt{3}{\mathcal{D}}_c^4}(3p_1-\rho_1)+\frac{1}{24}\left.\frac{dV}{d\phi}\right|_{\phi_c}$, where ${\mathcal{D}}_c\equiv {\mathcal{D}}(\phi_c)$.

Notice that, as $r\longrightarrow 0$, the Schwarzschild-like mass function approaches 0,
\be
    m(r)\approx\int_0^rd\Tilde{r}\left(\frac{4\pi}{{\mathcal{D}}_c^4}\Tilde{r}^2\rho_1+\frac{\Tilde{r}^2}{4}V(\phi_c)\right)\approx \left[\frac{4\pi}{3{\mathcal{D}}_c^4}\rho_1+\frac{1}{12}V(\phi_c)\right]r^3\ll r^2.
\ee

It is straightforward to compute $\delta_1(r)$ by direct substitution of \eqref{4} into \eqref{2} yielding
\be\label{delta1}
    \delta_1(r)=\frac{2\pi}{3{\mathcal{D}}_c^4}(p_o+\rho_1)(3p_1+\rho_1)r^2-\frac{1}{12}V(\phi_c)(p_1+\rho_1)r^2.
\ee

The initial conditions for $p_+(r)$ at the outer edge of the phase-2 nucleus, $r=r_n$, are

\be\label{IC_rn}
    p_+(r_n)=p_1,\qquad m(r_n)=\left[\frac{4\pi}{3{\mathcal{D}}_c^4}\rho_2+\frac{1}{12}V(\phi_c)\right]r_n^3.
\ee
Moreover, since $r_n$ is small,
\be\label{dp+}
    \frac{dp_+}{dr}(r_n)\approx -\frac{4\pi}{3{\mathcal{D}}_c^4}(3p_1+\rho_2)(p_1+\rho_1)r_n+\frac{1}{6}V(\phi_c)(p_1+\rho_1)r_n,
\ee
and
\be\label{delta_rn}
    \delta(r_n)\approx \frac{2\pi}{3{\mathcal{D}}_c^4}(3p_1+\rho_2)(p_1+\rho_2)r_n^2-\frac{1}{12}V(\phi_c)(p_1+\rho_2)r_n^2\ .
\ee

Substituting \eqref{3} in \eqref{2}, and neglecting terms proportional to $\Pi$ and $\Pi\frac{d\Pi}{dr}$, an equation analogous to the general relativistic one is left~\cite{Seidov}. Therefore, for small $r$, {the pressure difference} $\Pi(r)$ takes the form

\be\label{Pi}
    \Pi(r)=A+\frac{B}{r}.
\ee

Replacing \eqref{3} in \eqref{IC_rn},

\be
    A=\delta_1-\frac{B}{r_n},
\ee

\be
    \frac{B}{r_n^2}=-\left.\frac{d\delta_1}{dr}\right|_{r_n}-\left.\frac{dp_+}{dr}\right|_{r_n}.
\ee

By carrying out standard algebraic manipulations after solving the system and normalizing all quantities by the density on the lower end of the phase transition, $\rho_1$, one finds

\be\label{Adelta}
    \frac{A}{\delta}=\frac{1+\sigma_1}{q+\sigma_1}\frac{3\sigma_1+3-2q-K(\alpha)/\rho_1}{3\sigma_1+q-K(\alpha)/\rho_1},
\ee
where $\displaystyle\sigma_1=\frac{p_1}{\rho_1}$, $\displaystyle q=\frac{\rho_2}{\rho_1}$ and $K(\alpha)=\displaystyle\frac{3{\mathcal{D}}_c^4}{24\pi}V(\phi)$. Notice that we recover Seidov's limit in GR, Eq.~(\ref{Latent_heat2}), after taking the limit $K\longrightarrow 0$ in Eq.\eqref{Adelta},

\be\label{Adelta_GR}
    \left.\frac{A}{\delta}\right|_{\rm{GR}}=\frac{1+\sigma_1}{q+\sigma_1}\frac{3\sigma_1+3-2q}{3\sigma_1+q}.
\ee

Since $\displaystyle\frac{dM_+}{dp_c}=\frac{A}{\delta}\frac{dM}{dp_c}$, as in ~\cite{Seidov}, there exists a critical value $q_{crit}$ above which the derivative becomes negative and therefore the star is unstable and collapse. In GR this happens at 
\begin{equation}
q>q_{crit}=\frac{3}{2}(1+\sigma_1)\ ,
\end{equation}
{which is the well-known analytical Seidov limit}.

After several computer runs we can state that the central value of the dimensionless scalar field $\phi_c\sim 10^{-2} \ll 1$.
The central scalar field is obtained through a shooting method with a bisection process, which computes $\phi_c$ for a given $\alpha$. The algorithm works as follows: we first make an initial guess for $\phi_c$ and check whether the boundary conditions are satisfied. If they are not, the bisection method refines the guess until hitting an appropriate value of $\phi_c$. By applying this procedure to different configurations, we conclude that the central value of the dimensionless scalar field is $\phi_c\sim 10^{-2} \ll 1$.
Therefore, we can expand $K(\alpha)$ in modified gravity as

\be\label{K(alpha)}
    K(\alpha)\approx \frac{\phi_c^2}{24\pi\alpha}.
\ee

Keeping in mind that  $\rho_1\sim 10^{-3}\;\rm{km^2}$ and considering $\alpha\sim 10\;\rm{km^2}$, we arrive at

\be\label{x}
    x\equiv\frac{\phi_c^2}{24\pi\alpha\rho_1}\ll 1.
\ee

Then, we can write \eqref{Adelta} as

\be\label{Adelta2}
    \frac{A}{\delta}=\frac{1+\sigma_1}{q+\sigma_1}\frac{1}{3\sigma_1+q}\left[3\sigma_1+3-2q+\frac{3(1-q)}{3\sigma_1+q}x\right].
\ee

Finally, $q_{crit}$ in $R^2$-gravity is given by the following quadratic equation

\be\label{qcrit}
    3\sigma_1+3-2q+\frac{3(1-q)}{3\sigma_1+q}x=0.
\ee

It is easy to compute the critical value of $q$ that satisfies Eq.\eqref{qcrit}

\be\label{qcrit_sol}
    q_{crit}=\frac{3}{4}\left[1-\sigma_1-x\pm\frac{1}{\sqrt{3}}\sqrt{3+18\sigma_1+27\sigma_1^2+2x+6\sigma_1x+3x^2}\right].
\ee

Taking $x\longrightarrow 0$ we obtain two solutions, the general relativistic and an unphysical one. The positive solution takes an $R^2$-gravity correction given by the positive term in \eqref{qcrit_sol}. We discard the negative solution since it does not return Seidov's limit in GR and $q$ must be positive by definition. Finally, we can write \eqref{qcrit_sol} in a better looking way by expanding it around $x=0$ yielding 

\begin{eqnarray}\label{qcrit_sol2}
    q_{crit}& =& \frac{3}{2}\left(1+\sigma_1\right)+\frac{3}{4}\frac{1}{2(1+3\sigma_1)}\left[-\frac{4}{3}x-4\sigma_1x+x^2\right] \nonumber \\
    &=& q^{GR}_{crit}-\frac{x}{2}+o(x^2) \nonumber \\ &=& q^{GR}_{crit}+q_{crit}^{f(R)}[\alpha].
\end{eqnarray}

{Comparing with Eq.~(\ref{x}) we see that for positive $\alpha$ parameter the Seidov limit becomes larger in $f(R)$ theory. This feature in the small-core approximation persists, in numerical computations, for arbitrary sizes of the phase-2 nucleus. Thus, finding a star which exceeds the Seidov limit in GR + allowed EoS band immediately takes one to modified gravity.}

\section{Buchdahl-Bondi limit in $R^2$-gravity}
\label{sec:Bouchdahl-limit}
This limit is a maximum compactness which can be achieved in a neutron star: in General Relativity (for a static, spherically-symmetric star:  rotation could induce an EoS dependence), $R> \frac{9}{8}\times 2M$, nontrivially more stringent than the Schwarzschild limit $R>R_s=2M$.
It was extended in \cite{Goswami:2015dma} to $f(R)$ gravity, in which this Buchdahl-Bondi limit becomes

\begin{equation}\label{Bouchdahl}
    2M<\frac{4\frac{f_{,R}(R_0)}{f_{,R}(0)}\left(1+\frac{f_{,R}(R_0)}{f_{,R}(0)}\right)}{\left(1+2\frac{f_{,R}(R_0)}{f_{,R}(0)}\right)^2}R_s,
\end{equation}
where $R_0=R(0)$ is the Ricci scalar at the star's center. This means that the limit ceases to be universal, as the dependence on $R_0$ introduces not only the parameter $\alpha$ defining the theory of gravity, but also the equation of state of the matter content even in the static spherical-symmetric case. Since

\begin{equation}\label{R}
    R=\frac{2}{r^2}e^{-2\lambda}\left[-r^2\nu'^2+r^2\nu'\lambda'-r^2\nu''-2r\nu'+2r\lambda'+e^{2\lambda}-1\right]
\end{equation}
in a spherically symmetric and static spacetime with $ds^2=-e^{2\nu(r)}dt^2+e^{2\lambda(r)}dr^2+r^2(d\theta^2+\sin^2{\theta}d\phi^2)$, it follows from \eqref{Bouchdahl} ($f(R)=R+\alpha R^2$) that
\begin{equation}\label{exact_limit}
    M<\frac{4}{9}\frac{(1+2\alpha R_0)(1+\alpha R_0)}{\left(1+\frac{4}{3}\alpha R_0\right)^2}R_s.
\end{equation}

Notice that the general relativistic Buchdahl-Bondi limit is what remains of \eqref{exact_limit} upon taking $\alpha=0$. Furthermore, 
{in the opposite 
$\alpha\to\infty$ limit in Eq.\eqref{exact_limit}, in which $R^2$ is very dominant,}
\begin{equation}
    M<\frac{1}{2}R_s,
\end{equation}
we {curiously recover} the Schwarzschild limit. {Returning to theories around the GR case,} if we take $\alpha\ll 1$, {we pick up an order $\alpha^1$ correction,}
\begin{equation}
    M<\frac{4}{9}\left(1+\frac{1}{3}\alpha R_0\right)R_s.
\end{equation}

Recalling that $\lambda(0)=0$, {the last two terms of Eq.~(\ref{R}) drop out, leaving}
\begin{equation}\label{Ricci_scalar_0}
    R_0=\left.\frac{2}{r^2}\left[-r^2\nu'^2+r^2\nu'\lambda'-r^2\nu''-2r\nu'+2r\lambda'\right]\right|_{r=0}.
\end{equation}
{Evaluating it 
we can calculate \eqref{exact_limit} which we shall plot below}. 

{Here we can see again that, because $\nu'(0)$, $\lambda'(0)$ and $\nu''(0)$ depend} on the matter {content} through the field equations, {the Buchdahl-Bondi bound depends} on the central pressure {and should now be taken with an uncertainty band associated to the uncertainty in the EoS}. 

{As discussed in subsection~\ref{sec:field-eqs} (see Fig.~\ref{fig:R}), 
there are values of $\alpha$ that make the coefficient 
of $R_s$ in Eq.\eqref{exact_limit}
negative or divergent. 
This makes no sense and thus $\alpha$ must be carefully chosen.
The numerical computation will be presented later in section~\ref{sec:numerics}.}

\section{Stars in slow rotation to first order}
\label{sec:Rotating-star}

In this section we turn to a quick review of the theory for a slowly rotating star. Slow rotation is defined by the changes in pressure or energy density due to rotation to be small corrections. This implies that  particles at the surface of the star move only with non relativistic rotation velocities, i.e. $\Omega R\ll 1$, where $\Omega$ is the angular velocity of the surface of the star as seen by an observer at infinity. We will call $L(r,\theta)$ the angular velocity experienced by an observer in free fall towards the star, due to the dragging of the fluid. Then, we define $\varpi\equiv\Omega-L$ as the relative angular velocity. 

Following Hartle \& Thorne \cite{hartle:150,hartle:153}, the most general stationary axisymmetric metric takes the form
\begin{equation}\label{rot_metric}
    ds_*^2=-H^2dt^2+Q^2dr^2+r^2K^2(d\theta^2+\sin^2{\theta}(d\phi-Ldt)^2),
\end{equation}
where $H$, $Q$, $K$ and $L$ are functions of $r$ and $\theta$. The metric of this spacetime behaves in the same way under reversal in the direction of rotation as under a reversal in the direction of time. Due to this, an expansion of $H$, $Q$ and $K$ can only contain even powers of $\Omega$ whilst an expansion of $L$ can only contain odd powers of the angular velocity. Here we only consider terms to first order in $\Omega$ so that $L(r,\theta)=\omega(r,\theta)+O(\Omega^3)$ and the metric for the rotating star may be rewritten as
\begin{equation}\label{rot_metric2}
    ds_*^2=-e^{2\nu}dt^2+e^{2\lambda}dr^2+r^2(d\theta^2+\sin^2{\theta}d\phi^2)-2\omega dtd\phi,
\end{equation}

where $\omega(r,\theta)$ is linear in $\Omega$. In order to find the angular velocity we need to compute the field equation

\begin{equation}\label{Rt0}
    R_{\;\;\,\phi}^{*t}=8\pi T_{\;\;\,\phi}^{*t}.
\end{equation}

The 4-velocity of the rotating fluid is given by $u^{\mu}=(u^{t},0,0,\Omega u^{t})$. The normalization of $u$ gives, up to first order in $\Omega$, $u^{t}=e^{-\nu}$. The RH side of Eq.\eqref{Rt0} is easily computed from the perfect fluid's energy-momentum tensor. It is also straightforward to compute the LH side using the identity \cite{hartle:150}

\begin{equation}\label{Rt0_2}
    (-g^*)^{-1/2}R_{\;\;\,\phi}^{*t}=\partial_{\beta}\left[(-g^*)^{-1/2}g^{*t\alpha}\Gamma_{\phi\alpha}^{*\beta}\right].
\end{equation}

After some manipulations one finds the following equation for $\varpi(r,\theta)$,

\begin{equation}\label{varpi}
    \frac{e^{\nu-\lambda}}{r^4}\partial_r\left[e^{-\nu-\lambda}r^4\partial_r\varpi\right]+\frac{1}{r^2\sin^3{\theta}}\times\partial_{\theta}\left[\sin^3{\theta}\partial_{\theta}\varpi\right]=16\pi\mathcal{D}^{-4}(\rho+p)\varpi.
\end{equation}
We can now expand in Legendre polynomials so that
\begin{equation}\label{varpi_expansion}
    \varpi(r,\theta)=\sum_{l=1}^{\infty}\varpi_l(r)\frac{dP_l}{d\cos{\theta}},
\end{equation}
and substituting back into Eq.\eqref{varpi},  arrive at
\begin{small}
    \begin{equation}\label{varpi2}
    \frac{e^{\nu-\lambda}}{r^4}\partial_r\left[e^{-\nu-\lambda}r^4\frac{d\varpi_l}{dr}\right]+\frac{2-l(l+1)}{r^2}\varpi_l=16\pi\mathcal{D}^{-4}(\rho+p)\varpi_l.
    \end{equation}
\end{small}

The asymptotic exterior solution takes the form $\varpi\longrightarrow ar^{-l-2}+br^{l-1}$. Taking into account that $\displaystyle\varpi\longrightarrow \Omega-\frac{2J}{r^3}$, with $J$ the total angular momentum of the star, we can conclude that $l=1$ and therefore $\varpi_l$ vanish $\forall\;l\geq 2$. Thus, $\varpi_1\equiv\varpi(r)$ and the equation for $\varpi$ is

\begin{equation}\label{varpi3}
    \frac{e^{\nu-\lambda}}{r^4}\frac{d}{dr}\left[e^{-\nu-\lambda}r^4\frac{d\varpi(r)}{dr}\right]=16\pi\mathcal{D}^{-4}(\rho+p)\varpi(r),
\end{equation}

with boundary conditions $\displaystyle\lim_{r\to\infty}\varpi=\Omega$ and $\displaystyle\frac{d\varpi(0)}{dr}=0$. The first condition recovers the angular velocity as seen by the observer at infinity whilst the second condition guarantees the regularity at the center of the star. 

An observationally accesible quantity is the moment of inertia of the star, defined by
\begin{equation}\label{I}
    I=\frac{J}{\Omega}.
\end{equation}
Outside the star the term $\displaystyle e^{-\nu-\lambda}r^4\frac{d\varpi}{dr}$ in Eq.~\eqref{varpi3} is constant and has to match the interior solution at $r=R$. From this fact, an integral equation for the angular momentum of the star follows,
\begin{equation}\label{eq.J}
    \left.e^{-\nu-\lambda}r^4\frac{d\varpi}{dr}\right|_0^{R}=16\pi\int_0^{R}dr \mathcal{D}^{-4}(\rho+p)r^4e^{\lambda-\nu}\varpi=kJ.
\end{equation}
The constant $k$ is fixed by the Newtonian limit:
\begin{equation}\label{eq.Jnewt}
    J_{Newt}=\frac{8\pi}{3}\Omega\int_0^{R}dr (\rho+p)r^4.
\end{equation}

In the Newtonian limit $p\ll\rho$, $\phi(r)=0$ (and the conformal factor is $\mathcal{D}=1$), there is no dragging ($\omega=0$) and $\nu(r)=\lambda(r)$, so that $k=6$. Finally, we have obtained an equation for the moment of inertia

\begin{equation}
    I=\frac{8\pi}{3}\int_0^{R}dr\mathcal{D}^{-4}(\rho+p)r^4e^{-\nu+\lambda}\left(\frac{\varpi(r)}{\Omega}\right).
\end{equation}

This can be evaluated in modified gravity once the metric functions $\nu$, $\lambda$ have been determined with the help of the modified Einstein's equations, as we will do in the following section~\ref{sec:numerics}.

\section{Numerical computation}
\label{sec:numerics}

Let us now turn to a sketch of the numerical procedure and gather a few examples of the various physics points discussed. The key points are presented in Fig.~\ref{fig:Lmax}
and Fig.~\ref{fig:Delta}. The latter illustrates the correlation 
between the latent heat of any first-order phase transition and the discontinuity in a typical observable-to observable function, in this case moment of inertia as a function of mass; the latter, that the maximum latent heat is larger for modified gravity \textbf{and depends on its parameter $\alpha$. The Gravity Probe B experiment \cite{Naf:2010zy} provides that $|\alpha| \lesssim 5 \times 10^{15} \text{cm}^2$ therefore the obtained constraints on the parameter are consistent with the ones used further in the paper. Since we are exploring values of $\alpha$ in the range of 1-100 km$^2$, and the curvature scale is approximately (1/10 km)$^2$, our chosen values are slightly on the higher end compared to the constraints obtained from GW170817 in \cite{Jana:2018djs}.}

\subsection{Construction of the static star}
\begin{figure}[h!]
    \centering
\begin{tikzpicture}[node distance=2.5cm]

\node (start) [io] {Initial conditions: $\lambda_c$, $\nu_c$, $p_c$, $\rho_c$, $\phi_c$, $\phi'_c$};
\node (p1) [startstop, below of=start, yshift=0.5cm] {Runge-Kutta to solve Eq.(\ref{EF_Gtt})-(\ref{eq.phi22})};
\node (p2) [decision, below of=p1, yshift=-0.5cm] {Are BC's satisfied?};
\node (p3) [startstop, below of=p2, yshift=-0.5cm] {$\;\;\;$Solve the $\;\;\;$ ODE's system};
\node (p4) [io, below of=p3, yshift=0.5cm] {Save data in a file: e.g. $(M,R_s)$};

\node (p6) [startstop, right of=p2, xshift=2.8cm] {Bisection to find the new $\phi_c$};

\draw [arrow] (start) -- (p1);
\draw [arrow] (p1) -- (p2);
\draw [arrow] (p2) -- node[anchor=east] {yes} (p3);
\draw [arrow] (p3) -- (p4);
\draw [arrow] (p2) -- node[anchor=south] {no} (p6);
\draw [arrow] (p6) |- (p1);

\end{tikzpicture}
    \caption{Flowchart of the algorithm implemented in FORTRAN.}
    \label{fig:flowchart}
\end{figure}
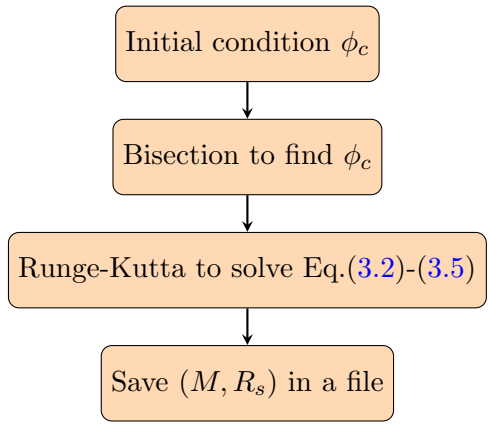

Since the EoS employed are computerized, we need to develop a numerical algorithm. We use a fourth order Runge-Kutta algorithm to address the differential TOV-like equations. The basic flowchart of the program\footnote{A sample of the code developed to compute the mass-radius diagrams can be found at \url{https://github.com/hyliano53/Modified-Gravity-NS.git}.} is shown in Fig.~\ref{fig:flowchart}.

A complication that arises is that we have boundary conditions, as we impose that the scalar field vanishes at infinity. Therefore, the Runge-Kutta algorithm is combined with a shooting method {(that employs a bisection) ensuring the boundary conditions.} Notice that the $\nu(r)$ function does not appear explicitly in any of the equations \eqref{EF_Gtt}-\eqref{eq.phi22} (only its derivative). Thus, we can solve the system for a given $\nu(0)=\nu_c$ and then obtain the appropriate function by subtracting {from  it a constant function equal to the value which $\nu$} takes at infinity $\nu(r)\longrightarrow\nu(r)-\nu(\infty)$ (and that continues to satisfy the differential equations).

The energy density $\rho(r_i)$ is obtained from the given pressure $p(r_i)$, through linear interpolation, allowing for integration on both sides of the phase transition. We also discuss the continuity of the solution in \cite{Moreno:2023xez} in the general-relativistic case, which could be extended to the present work.

Additional problems when numerically solving the system {arise because, as stated in \cite{Yazadjiev:2014cza},  the set of} differential equations is stiff, with increasing stiffness as $\alpha$ decreases. A poor guess for the initial condition of the scalar field during the shooting method makes $\phi(r)$, and consequently the other functions, eventually diverge. For small values of $\alpha$ it is much more difficult to obtain the desired $\phi_c$ due to the precision of the computer {(so recovering the GR results from $\alpha\to 0$ took quite some effort)}. 

In order to optimize the running time {the solution adopted consisted in 
setting a reasonably small interval around $r=0$ on which $\phi$ takes the desired $\phi_c$ value and is constant. Likewise, we  truncate the scalar field to 0 by hand at a distance at which the scalar field has decreased enough. The sensitivity of other quantities to these two grid cutoffs is examined to ensure
independence thereof.}

\begin{figure*}[t] 
  \centering
  \includegraphics[width=\textwidth]{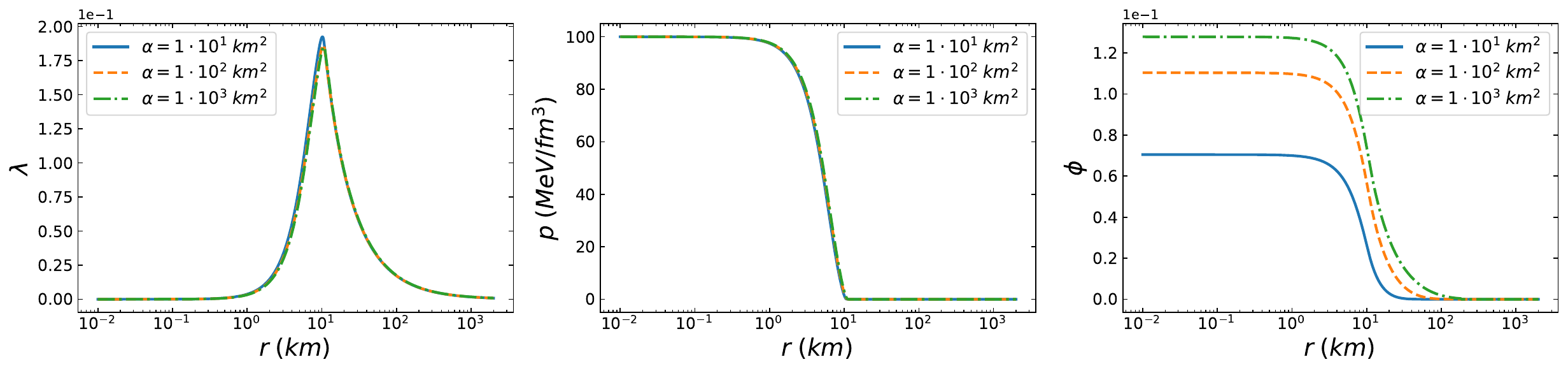}
  \caption{NS structure obtained computing the solution of the field equations for the EoS~``Interm'' shown in Fig.~\ref{fig:EoS}. 
{The profile is that of a star with} 100 $\rm{MeV/fm^3}$ central pressure. 
 {In the left panel (metric function $\lambda(r)$) and the middle one (pressure profile $p(r)$)}
  we can barely see {a dependence in $\alpha$}.
However, in the right panel, the scalar field profile $\phi(r)$ is clearly larger for the larger $\alpha$ values.}
  \label{fig:Stellar_functions}
\end{figure*}

We plot the $\lambda(r)$ function, the pressure profile and the scalar field in Fig.~\ref{fig:Stellar_functions}. As we can see, $\lambda$ reaches its maximum in the interior of the star. On the other hand, the pressure decreases monotonically until it vanishes at the edge of the star. In the last plot we can see again that the results are consistent with the boundary conditions, since the scalar field exponentially decreases outside the star, vanishing at large $r$.

The conclusion from the figure is that the structure of the matter in the NS itself is not very dependent on $\alpha$, as the metric function $\lambda$ and the pressure have profiles similar to those in GR. The scalar field however is of course very much $\alpha$-dependent, and hence the total mass and other overall properties of the possible stars within the modified gravity theory is different from GR.

\begin{figure*}[t] 
  \centering
  \includegraphics[width=\textwidth]{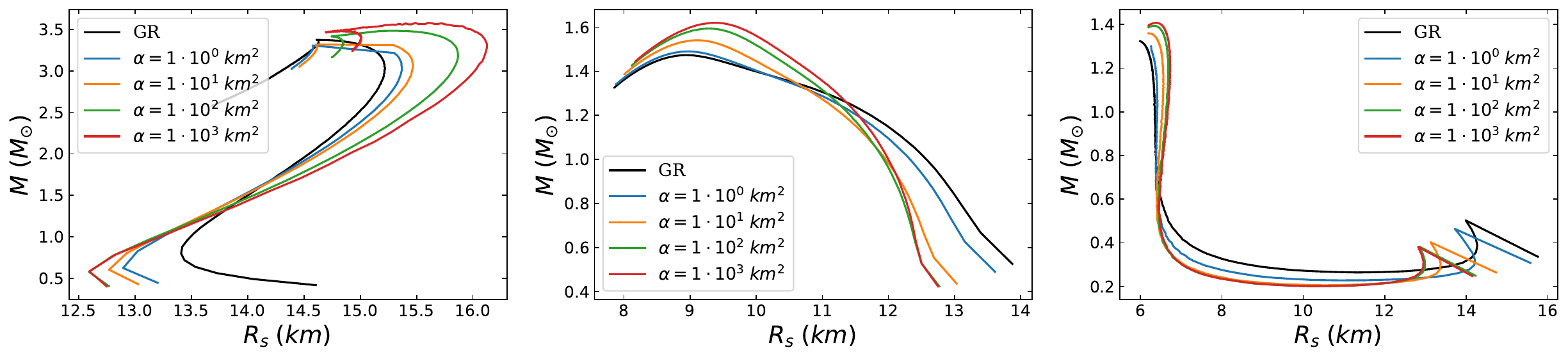}
  \caption{Mass-radius diagrams for different EoS from Fig \ref{fig:EoS} and several values of $\alpha$. Left panel: EoS ExR. Mid panel: EoS Interm (whitout phase transition). Right panel: EoS ExS}
  \label{fig:mass-radius}
\end{figure*}

This can be seen in Fig.~\ref{fig:mass-radius} where 
we plot the typical mass-radius diagram for different EoS and several values of the parameter of the theory. Notice that it is a common behaviour that the mass of the star grows with the $\alpha$ parameter for higher central pressures while it decreases with $\alpha$ for lower central pressures {(so the curves corresponding to different values of $\alpha$ cross)}. Furthermore, note that the limit of GR is recovered in the limit $\alpha\longrightarrow 0$ whilst the largest difference with GR is found for the greatest value of $\alpha$.

\subsubsection{The Buchdahl-Bondi limit}
\begin{figure}[h!]
\centering
\includegraphics[width=\textwidth]{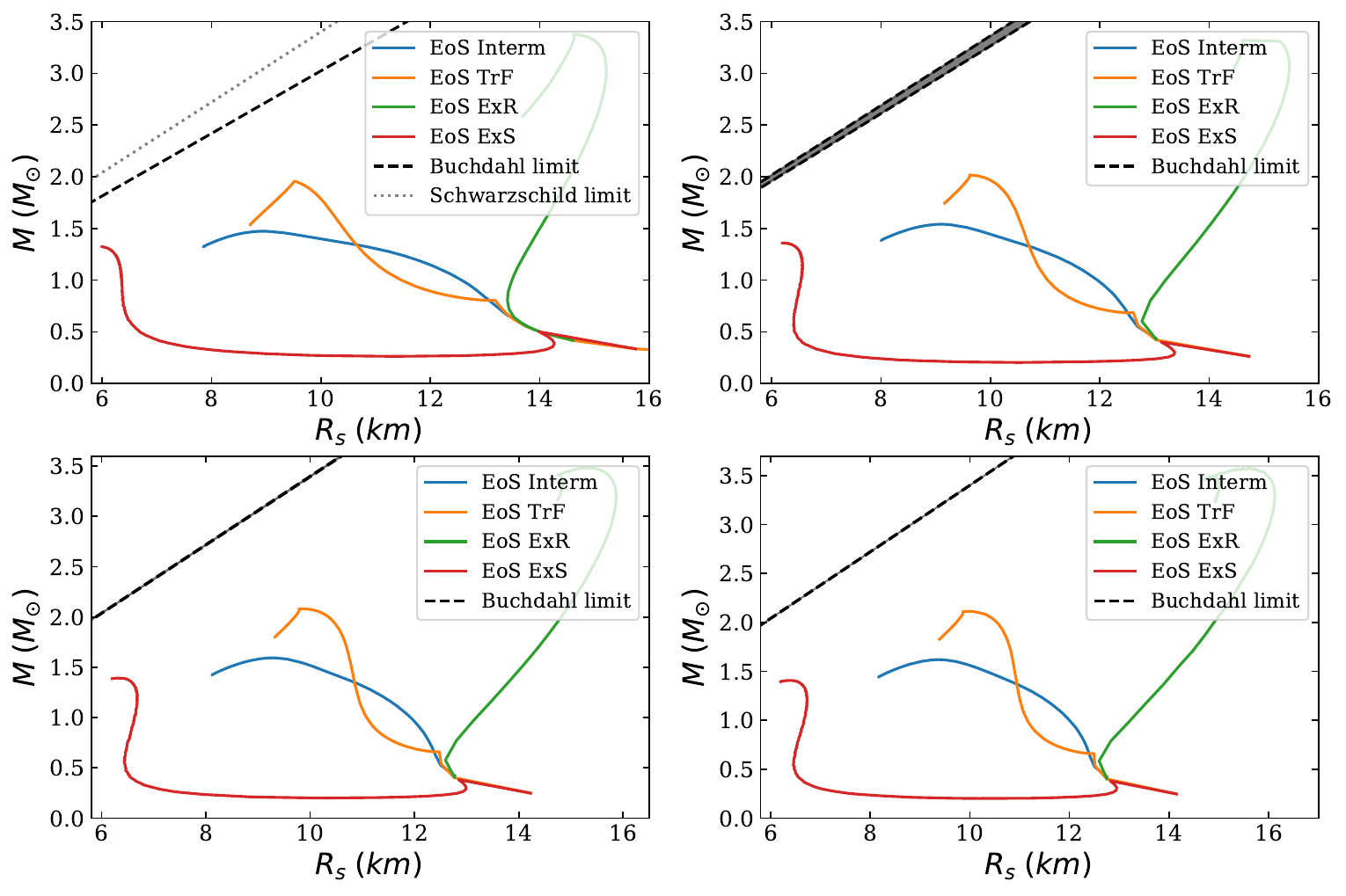}
\caption{ Buchdahl-Bondi limit (the diagonal black line on the top left corner of each panel). 
From top to bottom and left to right, the values of the $f(R)$ parameter are $\alpha=0$ (General Relativity), $\alpha=10\;\rm{km^2}$, $\alpha=1\cdot 10^2\;\rm{km^2}$ and $\alpha=1\cdot 10^3\;\rm{km^2}$.
A scatter of $M(R)$ lines for various (typical as well as extreme) EoS, in colour online, have been added for illustration.}
\label{fig:Buchdahl_limit}
\end{figure}

In Fig.~\ref{fig:Buchdahl_limit} we have added plots obtained computing the Buchdahl limit for different values of the parameter $\alpha$ of the theory. We have also used different EoS, including the extreme ones within the nEoS uncertainty band, to compute
an upper and a lower limit ({due to the dependence in $R_0$ which makes it sensitive to} the difference in the metric components derivatives for quite different central pressures).

For $\alpha=0$ (top-left panel in Fig.~\ref{fig:Buchdahl_limit} as well as for $\alpha$ very large
(bottom panels) there is no (or hardly) a difference between the two extremes of the Buchdahl-Bondi lines, so that the spread due to $R_0$ disappears. Of course, for large $\alpha$ both lines converge to the Schwarzschild limit $R_s=2M$ so the most interesting cases happen for intermediate $\alpha$ (top right panel).

\subsection{Nonanalyticities and latent heat}

Observing the various mass-radius diagrams displayed, we observe kinks (derivative discontinuities) which are due to the non-differentiability introduced by the phase transitions. For instance, we can see a kink near $(14.6\,\rm{km},3.5\,M_{\odot})$ in the EoS ExR GR diagram (and the analogous shifted kinks for $\alpha\neq 0$ curves). 

{These jumps in the derivative of the $M(R)$ function persist in modified gravity.}
For values of the parameter $\alpha$ greater than $10^3\;\rm{km^2}$ the mass-radius diagram does not significantly change anymore and thus the mass can hardly reach the typical 2 $M_{\odot}$ for the EoS in the mid and right panel of Fig.~\ref{fig:mass-radius}. The combination {of such high $\alpha$} with typical EoS, short of the stiffest ones, can thus be excluded.

We now come to one of the key results of this article. By {calculating all physically realizable stars with the} different EoS as in \cite{Lope-Oter:2024egz}, we obtain the maximum possible latent heat (the equivalent of the Seidov limit {but in modified gravity).}

\begin{figure}[h!]
\centering
\includegraphics[width=8cm]{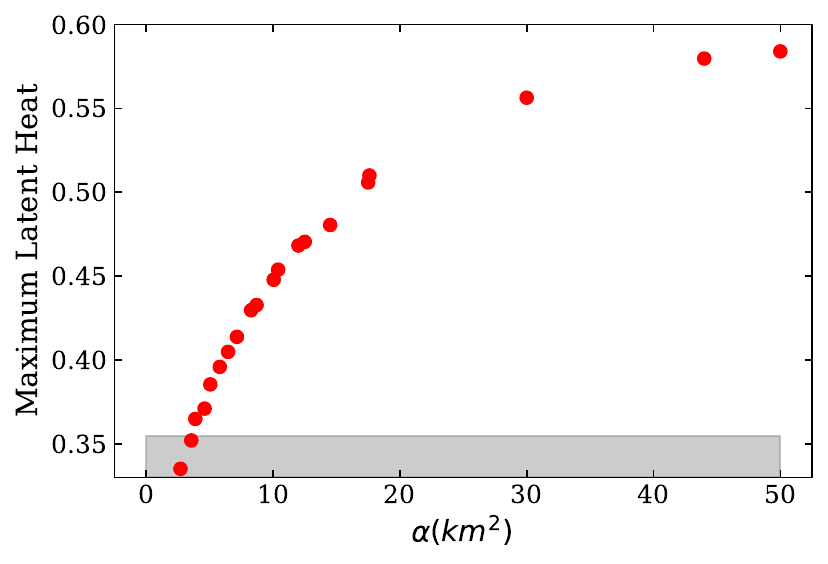}
\caption{Approximate maximum latent heat allowed within the EoS band of Fig.~\ref{fig:EoS}.
1) The points below the grey band (any with $\alpha<5$km$^2$) have a directly measurable Seidov limit which would be derived from gravitational collapse in $f(R)$ theory. 2) The latent heat of those with $\alpha$ between 5 and 50 km$^2$ will likely not be reachable because the star mass needed to produce the kink in a mass-radius diagram would be above 2.4$M_\odot$. 3) $L_{\rm max}$ above 0.58 is then forbidden by microscopic understanding of EoS uncertainty independently of the theory of gravity, so we stop plotting there.
}
\label{fig:Lmax}
\end{figure}

{We show in Fig.~\ref{fig:Lmax}
the dependence of this maximum latent heat (numerically computed) with the parameter $\alpha$ defining the modified gravity theory.
}

The maximum latent heats are reached for the stiffest EoS, in our band this is EoS ExR. 

We see in Fig.~\ref{fig:Lmax} that this maximum latent heat supported by the star increases with $\alpha$. 

This means that measuring a Seidov limit above the GR value in~\cite{Lope-Oter:2021mjp} would by necessity entail a violation of GR. The values of $\alpha$ manifestly accessible to this method would be those in $0<\alpha<5$km$^2$. For $\alpha$ above that, the star mass exceeds 2.4 solar masses and it is unclear that we would be able to measure such an extended phase transition, at least not in static stars. Fastly rotating stars and mergers might allow a higher $\alpha$ reach.

Finally, above $\alpha > 50$km$^2$, gravitational collapse ceases to be the tightest constraint imposed on $L$, since 
the allowed band of EoS in microscopic hadron physics limits $L$ (even for smaller neutron star masses). Thus, unless our understanding of the maximum latent heat from hadron physics is flawed, larger values of the parameter $\alpha$ do not produce singular behaviours for any of the EoS in the band. 

Only if nuclear physics would allow a greater uncertainty band than that of Fig.~\ref{fig:EoS}, could one eventually construct EoS with more extended phase transitions and could then search for a gravitational Seidov-like $L_{\rm{max}}$ limit also for $\alpha > 50$km$^2$.  

The conclusion is that latent heat is constrained by hadron physics for $\alpha>50$ and by gravitational collapse for $\alpha<50$ km$^2$, with the range below 5 accessible to static star studies, and that between 5 and 50 requiring higher pressures and thus more dynamical information.

Almost all of the points shown lie above the gray band, which marks the EoS with the exact phase transition that leads to a 2.4$M_\odot$ maximum mass. Thus, the $f(R)$ Seidov limit is not reachable. Therefore, if the one in GR is broken, only a lower bound on $\alpha$ could be extracted, which would still provide valuable information.

\subsection{Slowly rotating star}
Here we finally turn to the 
nonanalyticity of the moment of inertia which can be exposed in observations of both angular frequency and angular momentum (for example, with a third generation gravitational wave detector~\cite{Gupta:2024bqn,Maggiore:2019uih}) and its dependence on the intensity of the modification of gravity.

\begin{figure}[h!]
\centering
\includegraphics[width=8cm]{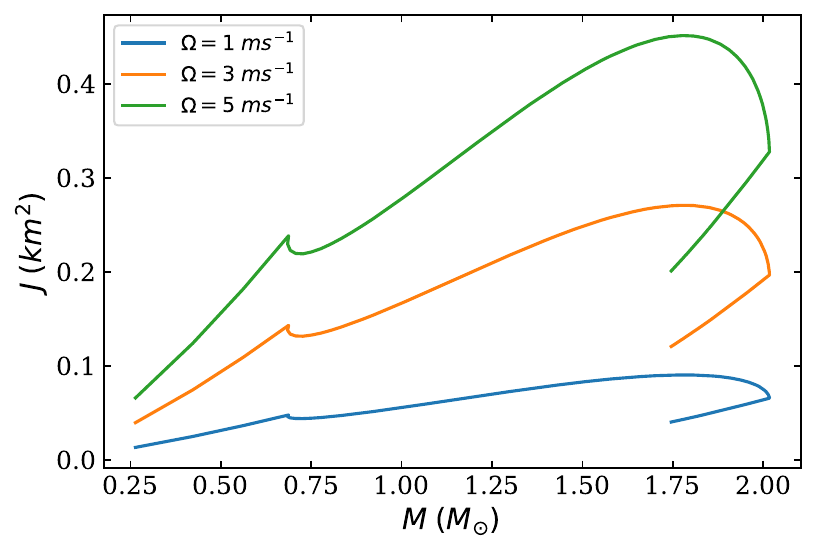}
\caption{Angular momentum of a family of stars (the one with EoS TrF) in $R+\alpha R^2$ theory with $\alpha=10\;\rm{km^2}$. Each curve has a different angular velocity as indicated (colour online).}
\label{fig:J_EoS_TrF}
\end{figure}

In Fig.~\ref{fig:J_EoS_TrF} we show the total angular momentum of a family of stars as a function of their mass for different angular velocities. Notice the kink around 0.65 $M_{\odot}$ due to the first order phase transition present in this particular example EoS TrF (orange line in Fig.~\ref{fig:EoS}).

\begin{figure}[h!]
\centering
\includegraphics[width=8cm]{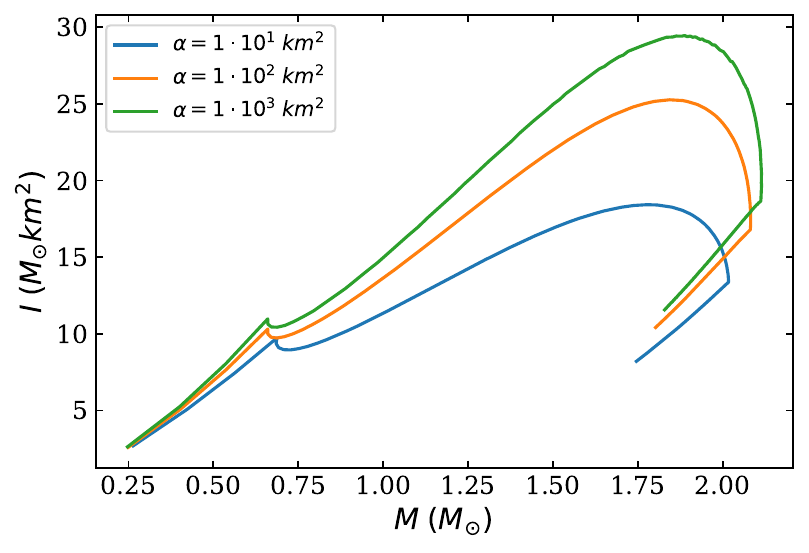}
\caption{Moment of inertia of a family of stars (EoS TrF) {all rotating with } the same angular velocity $\Omega=1\;\rm{ms^{-1}}$. Different $\alpha$ values for as shown in the legend (colour online).}
\label{fig:I_EoS_TrF}
\end{figure}

\begin{figure}[h!]
\centering
\includegraphics[width=\textwidth]{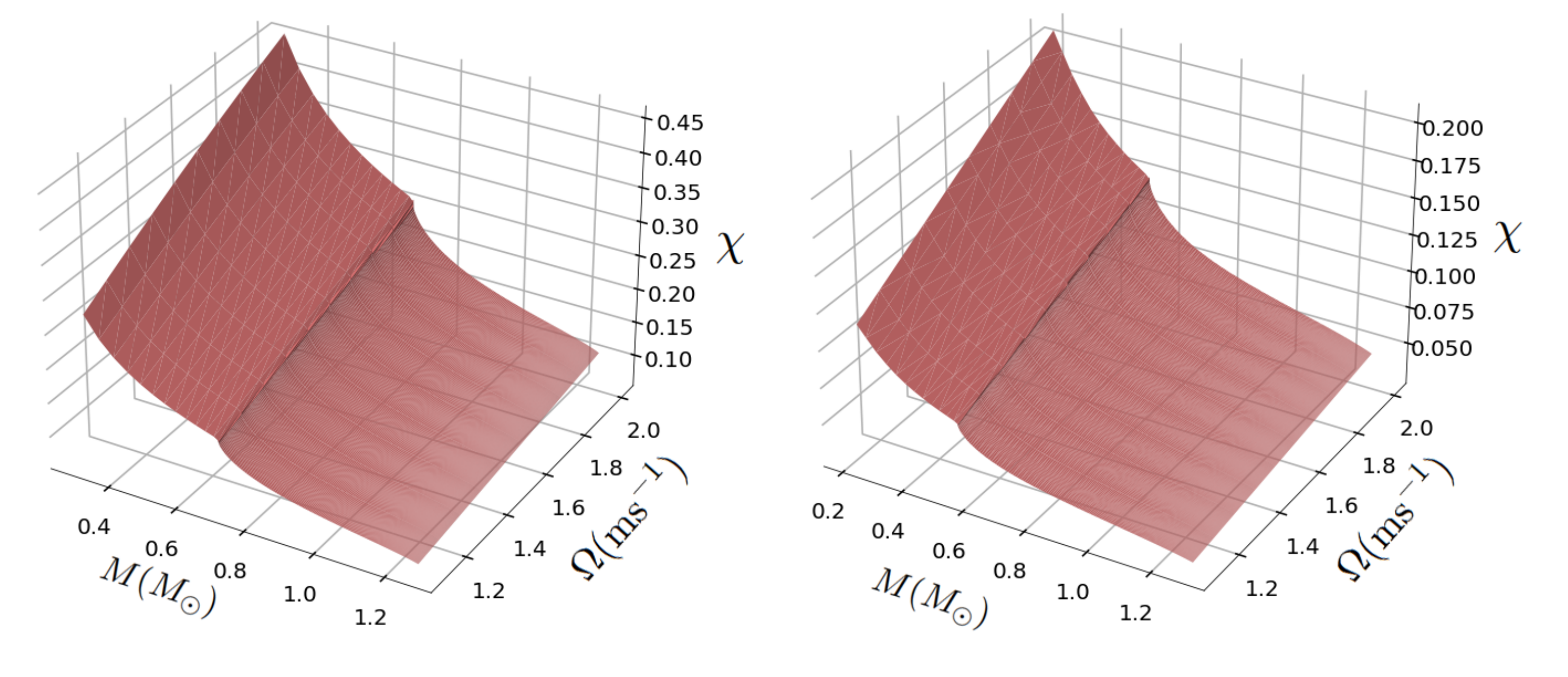}
\caption{Adimensional angular momentum $\chi$ as a function of the mass $M$ and the angular velocity of the star $\Omega$ for EoS TrF 3. Left panel: result in GR. Right panel: result in $R^2-$gravity with $\alpha =10^3\;\rm{km^2}$. {(The two plots look similar, but notice the different dimensionless spin $\chi$ scale. The ridge is also displaced at a different stellar mass.)}}
\label{fig:chi}
\end{figure}

In Fig.~\ref{fig:I_EoS_TrF} we then display the moment of inertia of {the same family} of stars as a function of their mass. Again, we can see a clear non-analyticity due to the phase transition present in EoS TrF. According to our previous results \cite{Moreno:2023xez}, the angular momentum of the star increases with the angular velocity and the same happens for the moment of inertia. In Fig.~\ref{fig:chi} we show the adimensional angular momentum $\chi$, defined as $\chi=J/M^2$, accessible for example through gravitational radiation in binary mergers \cite{LIGOScientific:2020kqk,LIGOScientific:2018hze}. We can also see a clear ridge due to the phase transition.

We aim to extract the intensity of these kinks from experimental data and correlate them, through theoretical models, with the latent heat of neutron star matter. To be precise, we have calculated the discontinuity in the derivative of the moment of inertia with respect to mass.

This discontinuity is defined as the difference in $\frac{dI}{dM}$ across the two sides of the kink. Fig.~\ref{fig:EoS} illustrates this calculation, where we employed similar EoS with phase transitions at intermediate pressures (EoS TrF 2-5), enabling a comparison of $\frac{dI}{dM}$ between them.

\begin{figure}[h!]
\centering
\includegraphics[width=8cm]{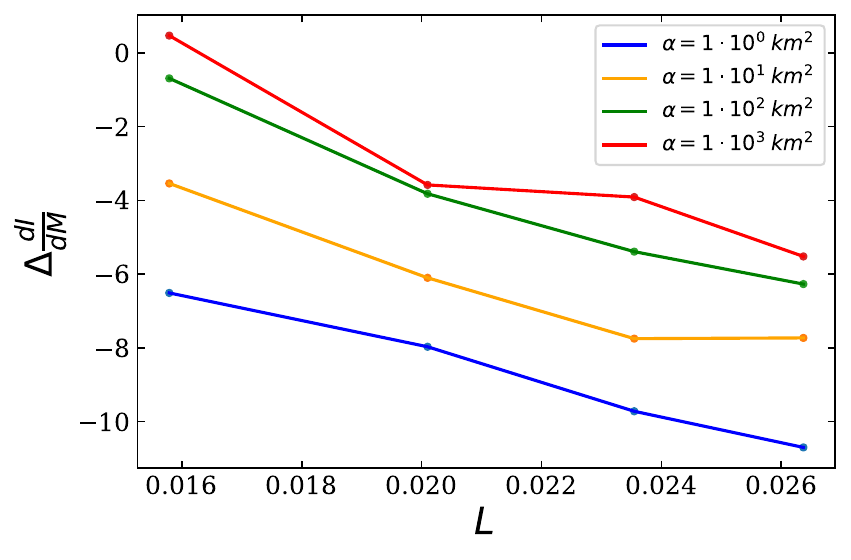}
\caption{$\Delta\frac{dI}{dM}$ as a function of the latent heat. Similar EoS from Fig.~\ref{fig:EoS} employed due to the strong dependence of $I(M)$ with the EoS, leading to very different shapes for distinct EoS.}
\label{fig:Delta}
\end{figure}

To clarify what can be extracted from observations, we have computed the jump in the derivative $\Delta\frac{dI}{dM}$ at $\pm 0.1M_{\odot}$ around the kink's position in Fig.~\ref{fig:Delta}. This approach is also inspired by Lindblom's analysis of the mass-radius diagram \cite{Lindblom:1998dp}, which shows that taking the derivative arbitrarily close to the phase transition point is ineffective. The $M(R)$ curve has a technically continuous derivative in the cases he examined, and only over a finite interval does the drastic change in the direction of the $M(R)$ tangent become evident.

From the plot, we observe that $\left|\Delta\frac{dI}{dM}\right|$ decreases with increasing $\alpha$, while it grows with the latent heat $L$, indicating that a stronger phase transition amplifies the slope difference after the kink. This behavior clearly reflects how the phase transition intensity influences the moment of inertia's derivative.

Additionally, if one were to compute the field equations up to second order in angular velocity (which is beyond the current scope for $f(R)$ theories), it would be possible to obtain the mass and radius corrections. These corrections could be used to calculate the ellipticity, providing insight into the star’s deformation due to rotation \cite{Moreno:2023xez,hartle:150,hartle:153}.



\section{Conclusions}
\label{sec:Conclusions}

In this work we have studied neutron star properties in modified gravity, {following by now standard procedures}. First, we have reviewed a general formalism in order to write the Tolman-Oppenheimer-Volkoff equations for a quite general family of theories. Then we have studied $f(R)$ theories and their equivalence to scalar-tensor theories, computing a complicated TOV-like system of differential equations that must be solved numerically. In particular, we studied $R^2$-gravity following \cite{Staykov:2016mbt,Yazadjiev:2014cza}. Next, we presented the set of equations of state used along this work. The EoS employed have interesting properties due to the first order phase transitions and the fact that they are not constrained by any astrophysical observable, making them useful for our study in modified gravity \cite{Lope-Oter:2023urz,Lope-Oter:2021vxl}. 

We have solved the static star and computed the mass-radius diagram for several families of stars with different EoS within the allowed band thereof. We reiterate that there are points of non analyticity (or ridges in multidimensional plots) due to any first order phase transitions, and observed how they change with the deviation from GR, that is, for a few values of the parameter $\alpha$.

Finally, we have studied the slowly rotating star, which is a good approximation for most known pulsars. We computed the angular velocity equation in modified gravity in the same way as H\&T did in general relativity \cite{hartle:150,hartle:153}. We calculated some physical observables, such as the moment of inertia, which could be measured by future experiments. We studied how the angular momentum of the family of stars changes with the angular velocity and showed that the first order phase transitions present in the EoS leaves a clear kink in the $J(M)$ and $I(M)$ diagrams. We also showed a ridge in the $\chi(M,\Omega)$ diagram due to the non-analyticity in the EoS. We concluded {that exercise by} studying how the moment of inertia changes with the parameter of the theory. Again, future observations could constrain the free parameter $\alpha$ of the theory comparing with these numerical calculations.

This would {in principle} allow experimental measurements to eventually observe how the theory drifts away from GR.
To do it with these method, the observable-to-observable diagrams will have to be populated with multiple well-measured stars; this identification of a phase transition and eventual modification of GR in population studies is distinct from direct detection of a frequency-peak shift in one or few events which has also been suggested as a strong tell-tale signal~\cite{Bauswein:2018bma} of exotic phenomena in neutron stars.

{A difficulty to carry out that program is that any 
derivative jump found could also be caused by a gravitational phase transition. As has been reported~\cite{Doneva:2023kkz} in the literature, this phenomenon arises when a dynamical threshold, for example a certain energy density or a certain value of the curvature scalar, is reached. Then, a field might switch on from zero to a finite value, causing a point of nonderivability in star properties. This causes an ambiguity with phase transitions in the ordinary neutron matter of the star. An observation exceeding the Seidov limit helps to lift the degeneracy by necessarily pointing to modified gravity as no matter content can break it in General Relativity.}

{To overcome the difficulty,}  we have computed the maximum latent heat as a function of the parameter of the theory following the idea of Seidov in GR \cite{Seidov}. 

{ In the future, a combination of observations which populate observable-to-observable diagrams could identify a sudden derivative change (for example, in a diagram involving the stellar mass, we propose to use a $\pm 0.1 M_\odot$ interval as a reasonable criterion), which would correlate with a latent heat and test the Seidov limit. Its breach would lift the matter/gravity degeneracy and guarantee that we are witnessing a modified gravity phenomenon, be it because of a gravitational phase transition, be it because of a matter phase transition which should not be observable in GR.   
}

{We have also studied the Buchdahl-Bondi limit in $f(R)$ modified gravity following earlier authors, note that the limit becomes blurred into a band, acquiring a dependence on the EoS, and find it, for the time being, less promising than the Seidov limit for modified gravity searches.}

Although we have limited ourselves to cold stars, an extension to finite temperature is straightforward and our results should be easy to map, as the finite temperature neutron star EoS has been well discussed in the literature~\cite{Routray:2024kgv,Lope-Oter:2021vxl,Kochankovski:2022woy}.

A promising future additional research direction 
to lift the gravity/matter degeneracy and investigate whether a microscopic phase transition can be distinguished from a gravitational one proceeds by the search for twin stars.
Most EoS models are predominantly hadronic, but {should there be} an unconfined-quark core~\cite{Anand:1980ye,Thakur:2022sty}, or other exotic matter, which may exist at the center of neutron stars, this could give rise to higher-density  compact stars or ``hybrid stars''~\cite{Rosenhauer:1992vx}. Such phase transitions can cause the discontinuities {which we have been discussing. Focusing on}  the mass-radius relationship, this can lead  to the formation of a second branch of solutions, with the same mass but  smaller radius than neutron stars from the original branch, known as ``twin stars''—two neutron stars with equal mass but different radii ~\cite{Bhattacharyya:2004fn}.

In general relativity, twin stars fall into four categories based on the relationship between the maximum masses of neutron stars~\cite{Alford:2015dpa}. However, modified gravity introduces a fifth category (denoted as Ia in the literature) that does not exist in general relativity \cite{Lope-Oter:2024egz}. This addition creates ambiguity, so we should be cautious in claiming that future radius measurements from the NICER mission could confirm the existence of strong phase transitions in dense neutron star matter by identifying twin stars. {Our group is researching this observable to try to lift the matter/gravity degeneracy also here}.

\acknowledgments
We thank Eva L. Oter for providing her EoS set and for discussions.\\
Work supported by grant PID2022-137003NB-I00 and PID2022-138607NB-I00 of the Spanish MCIN/AEI
/10.13039/501100011033/ and PRX23/00225 (estancias en el extranjero); EU’s ERDF A way of making Europe and 824093 (STRONG2020); and Univ. Complutense de Madrid under research group 910309 and IPARCOS- UCM/2023 graduate assistance program. AW acknowledges financial support from MICINN (Spain) {\it Ayuda Juan de la Cierva - incorporaci\'on} 2020 N$^{\rm o}$. IJC2020-044751-I.



\end{document}